\pdfoutput=1
\RequirePackage[svgnames,table]{xcolor}

\documentclass[11pt,letterpaper]{yalearxiv}

\usepackage{graphicx}
\usepackage{float,epstopdf}
\usepackage{bbm}

\usepackage{microtype}

\usepackage{natbib}
\setcitestyle{square}

\usepackage{subcaption}
\usepackage{booktabs}
\usepackage{longtable}
\usepackage{nicefrac}

\usepackage{amsmath}
\usepackage{amssymb}
\usepackage{mathtools}
\usepackage{amsthm}
\usepackage{dsfont}
\usepackage{multicol}
\usepackage{makecell}
\usepackage{multirow} 
\usepackage{amsfonts} 
\usepackage{mathrsfs}
\usepackage[amssymb, thickqspace]{SIunits}
\usepackage{enumitem}
\usepackage{pgfplotstable}
\pgfplotsset{compat=1.18}
\usepackage{lipsum}		

\usepackage{microtype}
\usepackage{graphicx}
\usepackage[table]{xcolor}
\usepackage{arydshln}
\usepackage[normalem]{ulem} 

\usepackage{cases}
\usepackage{wrapfig}

\usepackage{url}

\usepackage{thmtools}
\usepackage{thm-restate}
\usepackage{tabu}

\definecolor{huskypurple}{HTML}{4B2E83}

\usepackage{titletoc}

\usepackage{listings}
\lstdefinestyle{promptstyle}{
  basicstyle=\ttfamily\footnotesize,
  breaklines=true,
  breakautoindent=false,
  breakindent=0pt,
  postbreak=\mbox{\textcolor{gray}{$\hookrightarrow$}\space},
  columns=fullflexible,
  keepspaces=true,
  frame=single,
  framesep=5pt,
  xleftmargin=6pt,
  xrightmargin=6pt,
  aboveskip=8pt,
  belowskip=8pt,
  showstringspaces=false,
}   
\makeatletter
\def\munderbar#1{\underline{\sbox\tw@{$#1$}\dp\tw@\z@\box\tw@}}
\makeatother

\AddToHook{cmd/appendix/before}{%
  \setcounter{axiom}{0}%
}

\newcommand{\be}{\begin{equation}}
\newcommand{\ee}{\end{equation}}
\newcommand{\bea}{\begin{equation*}\begin{aligned}}
\newcommand{\eea}{\end{aligned}\end{equation*}}

\usepackage{fontawesome5}

\newtcolorbox{payloadbox}{
    colback=infinibg,
    colframe=YaleLightBlue!45,
    boxrule=0.6pt,
    arc=2mm,
    left=10pt, right=10pt,
    top=6pt, bottom=6pt,
    before skip=5pt,
    after skip=5pt,
    fontupper=\footnotesize\ttfamily\raggedright,
}

\newtcolorbox{appendixindexbox}{
    colback=infinibg,
    colframe=YaleLightBlue!35,
    boxrule=0.5pt,
    arc=2mm,
    left=12pt, right=12pt,
    top=10pt, bottom=10pt,
    before skip=4pt,
    after skip=14pt,
}

\title{The Surface You Test Is Not the Surface That Breaks}

\runningtitle{The Surface You Test Is Not the Surface That Breaks}

\papervenue{Accepted at NeurIPS 2026 Workshop (AIWILD)}

\correspondingauthor{\textsuperscript{*}Corresponding author: \texttt{shifat.arman@eng.ox.ac.uk}.}

\author{%
  Syed Nazmus Sakib\textsuperscript{1},\;
  Nafiul Haque\textsuperscript{1},\;
  Shahrear Bin Amin\textsuperscript{1},\;
  Shifat E. Arman\textsuperscript{1,2,*}\\
  {\normalfont\fontsize{10}{13.5}\selectfont
    \rule{0pt}{1.45em}\textsuperscript{1}Department of Robotics and Mechatronics Engineering, University of Dhaka, Bangladesh\\
    \textsuperscript{2}Department of Engineering Science, University of Oxford, UK\\
    \rule{0pt}{1.5em}\href{https://syed-nazmus-sakib.github.io/CrossSurface}{\faGlobe~Project Page}%
    \qquad\href{https://github.com/syed-nazmus-sakib/surface-adaptive-injection}{\faGithub~Code}%
  }%
}

\hypersetup{
    colorlinks=true,
    linkcolor=YaleLightBlue,
    citecolor=YaleLightBlue,
    urlcolor=YaleLightBlue,
}

\begin{document}

\begin{abstract}
  Prompt-injection benchmarks for LLM agents typically test attacks through a single injection surface and report the resulting attack success rate as a property of the model. We ask whether those robustness conclusions remain stable when the same adversarial content enters through a different part of the agent interface. Using AgentDojo, we evaluate 13 LLMs across four task suites and place a byte-identical payload either in a tool output or in the tool description. This small change produces large differences in comparative robustness: 44.9\% of all model pairs change their relative ordering across the two surfaces, with substantial ranking instability in every suite. The effect is especially pronounced for a small number of models, showing that a benchmark can substantially underestimate vulnerability when it tests only one surface. We further find that this behavior has predictive structure. Using three suites to identify the riskier surface for each model predicts the more vulnerable surface on an unseen suite with 76.9\% accuracy. Defense results show the same dependence, as mitigations effective against tool-output attacks can leave substantial exposure through tool descriptions. Our results show that prompt-injection robustness is not surface-invariant and that agent evaluations should test the surfaces on which their security conclusions depend.
\end{abstract}

\maketitle

\section{Introduction}

Tool-augmented language-model agents are increasingly used in settings where they interact with external tools, APIs, retrieved documents, and other potentially untrusted sources of information \citep{schick2023toolformer,patil2024gorilla,qin2024toolllm}. These capabilities introduce a fundamental security problem because adversarial instructions can enter an agent's context through content that does not originate from the user and can cause the agent to execute unintended actions \citep{greshake2023not}. Existing benchmarks such as AgentDojo and InjecAgent have made this threat measurable by injecting adversarial content into tool-mediated interactions and reporting an attack success rate for each model \citep{debenedetti2024agentdojo,zhan2024injecagent}. Such evaluations are useful, but they often test a particular injection surface and implicitly treat the resulting vulnerability score as a property of the model. This interpretation assumes that a model judged safer than another under one attack surface would remain safer when the same adversarial instruction arrives through a different part of the agent's context.

That assumption is increasingly difficult to justify. Tool outputs are a well-studied source of indirect prompt injection, but tool descriptions and other parts of the tool schema are also established attack surfaces \citep{shi2025prompt,zhang2025allies,wang2026mcptox}. System prompts, retrieved content, and multimodal inputs provide additional channels through which adversarial instructions can reach an agent \citep{greshake2023not,bagdasaryan2023abusing,li2025system}. The important question is therefore no longer whether multiple attack surfaces exist. We instead ask whether \textbf{the security conclusions produced by an agent evaluator remain stable when the evaluated surface changes}. If the same adversarial content can make one model appear safer on one surface and less safe on another, then a single-surface evaluation can change not only the measured level of risk but also the comparative judgment about which model is more robust.

We study this question through a controlled cross-surface evaluation built on AgentDojo. We evaluate 13 LLMs across the banking, Slack, travel, and workspace suites, giving 52 model-suite cells. For every matched user-task and injection-task pair, we place the same byte-identical adversarial payload either in the return value of an attacker-controlled tool, which we call the data surface, or in the natural-language description of that tool, which we call the schema surface. The task, payload, agent harness, and success predicate remain unchanged, isolating the effect of where the adversarial content enters the agent. We find that security rankings are highly sensitive to this choice. Changing from data-surface to schema-surface evaluation reverses the ordering of 35 of 78 model pairs, or 44.9\%, with inversion rates ranging from 38.5\% to 57.7\% across individual suites. The underlying differences can also be large: Gemini-3-Flash averages 8.8\% attack success on the data surface and 74.6\% on the schema surface, while GPT-4.1 exhibits the opposite pattern with 54.9\% and 6.8\%, respectively. A verifier restricted to one surface can therefore reach a substantially different conclusion about which model is safer even though the malicious instruction itself is unchanged.

This instability also has predictive structure. In a leave-one-suite-out experiment, we use three suites to identify the more vulnerable surface for each model and apply that choice to the unseen fourth suite. The resulting verifier selects the riskier held-out surface with 76.9\% accuracy and reaches 44.0\% attack success, compared with 37.5\% under data-only evaluation and 46.5\% for an oracle that observes both surfaces. It therefore recovers roughly 72\% of the oracle improvement without using target-domain attacks to choose the surface. The absolute underestimation is nevertheless highly uneven across models: the mean worst-surface increase over data-only evaluation is 9.1 percentage points, while the median is only 1.4 points, with much of the mean driven by Gemini-3-Flash and GPT-5.4. Our claim is therefore not that every model hides a large vulnerability, but that comparative security conclusions are not invariant to the surface on which an agent is evaluated.

The same issue appears when defenses are evaluated. Prompt-level defenses such as repeating the user instruction or spotlighting substantially reduce attack success on tool outputs but leave much higher residual vulnerability on tool descriptions, whereas a description-aware tool filter suppresses both tested surfaces. This suggests that security evaluation should treat vulnerability as a property of the interaction between a model and an attack surface rather than as a single scalar associated with the model. We make three contributions. We first provide a byte-identical cross-surface evaluation across 13 LLMs and four AgentDojo suites. We then quantify verification instability through pairwise ranking inversions, worst-surface underestimation, and held-out-domain surface selection. Finally, we show that surface incompleteness extends to defense evaluation and argue that comparative robustness claims should be tested across the realistic surfaces available to an attacker.

\section{Related Work}

\paragraph{Prompt injection in tool-augmented agents.}
Indirect prompt injection exploits the inability of language models to reliably separate instructions from untrusted content that enters their context \citep{perez2022ignore,greshake2023not}. As LLMs have become increasingly integrated with external tools, this threat has expanded from malicious retrieved text to attacks that manipulate an agent's tool-mediated observations and actions. InjecAgent evaluates indirect prompt injection in tool-integrated agents across diverse application scenarios, while AgentDojo provides a stateful benchmark in which agents must simultaneously complete legitimate user tasks and resist adversarial instructions delivered through their environment \citep{zhan2024injecagent,debenedetti2024agentdojo}. Subsequent work has developed stronger attacks and defenses within these settings, including adaptive attacks that explicitly optimize against existing mitigations \citep{yi2025benchmarking,zhan2025adaptive,nasr2026attacker}, while recent evaluation frameworks broaden comparison across attacks, defenses, applications, tasks, and benchmarks \citep{cui-etal-2026-rethinking,geng2026piarena}. These benchmarks have made attack success rate a standard measure of agent security, but the reported score is generally conditioned on the particular channel through which the benchmark introduces adversarial content. Our work studies a complementary question and asks whether the comparative security conclusions obtained from such evaluations remain stable when the evaluated surface itself changes.

\paragraph{Tool poisoning and schema-level attacks.}
Tool outputs are not the only untrusted component exposed to an agent. ToolHijacker and ToolCommander demonstrate that malicious or manipulated tool descriptions can influence tool selection and induce unintended behavior in tool-using language models \citep{shi2025prompt,zhang2025allies}. MCPTox further studies tool poisoning in real-world Model Context Protocol servers and explicitly considers malicious information embedded in tool metadata \citep{wang2026mcptox}. Full-Schema Poisoning broadens this threat model beyond natural-language descriptions by showing that adversarial content can be placed throughout structured tool schemas, while related work has demonstrated additional injection channels at the system-prompt and multimodal levels \citep{li2025system,bagdasaryan2023abusing}. We therefore do not treat tool descriptions as a previously unidentified attack surface. Instead, our distinction is experimental: we hold the adversarial payload byte-identical and vary only whether it appears in a tool output or a tool description, allowing us to measure whether the \textbf{security judgment produced by the evaluator changes with the surface being tested}. This controlled comparison is important because prior schema-poisoning work primarily establishes that a surface can be attacked or develops attacks specialized to that surface, whereas our focus is the stability of model comparisons across surfaces.

\paragraph{Adversarial evaluation and defense completeness.}
A long line of work in adversarial machine learning has shown that robustness claims can depend strongly on the attacks included in an evaluation and that fixed or non-adaptive threat models can substantially overestimate security \citep{athalye2018obfuscated,carlini2019evaluating,tramer2020adaptive,croce2020reliable}. This principle has subsequently shaped the evaluation of language-model jailbreaks and prompt injection, where stronger or adaptive adversaries frequently expose vulnerabilities that are hidden under a single attack configuration \citep{zou2023universal,chao2025jailbreaking,andriushchenko2025jailbreaking,zhan2025adaptive}. Prompt-injection defenses similarly range from techniques that mark or isolate untrusted content, such as Spotlighting, to approaches that train models to prioritize instructions according to their authority, such as Instruction Hierarchy \citep{hines2024defending,wallace2024instruction}. Our work extends the robustness-evaluation perspective from variation in attack content or optimization strategy to variation in \textbf{where the same attack enters the agent}. Rather than proposing a new attack algorithm, we examine whether a verifier restricted to one surface preserves model rankings, accurately estimates worst-surface risk, and reaches conclusions that generalize across task domains. We also apply the same perspective to defenses by asking whether low residual vulnerability on the surface used for evaluation remains low when the attack is delivered through another established channel. Accordingly, we do not claim that existing defenses are uniformly ``surface-blind'': the description-aware \texttt{tool\_filter} defense suppresses both tested surfaces, and our results instead show that this asymmetry is specific to prompt-level defenses.

\section{Experimental Design}

\subsection{Evaluation environment and model panel}

We conduct our primary evaluation using \textbf{AgentDojo} \citep{debenedetti2024agentdojo}, a stateful benchmark for studying prompt-injection attacks against tool-using language-model agents. AgentDojo contains four task suites, banking, Slack, travel, and workspace, each comprising legitimate user tasks and adversarial injection tasks that can be executed through the environment. An evaluation episode pairs a user task $u$ with an injection task $i$, after which the agent interacts with the available tools while attempting to complete the user request. Attack success is determined from the resulting environment state and tool-call trace using the benchmark's task-specific predicates.

We evaluate 13 LLMs from six providers. Every available model is evaluated on the same four suites, producing 52 model-suite cells. All models are accessed through OpenRouter (\url{https://openrouter.ai}) and evaluated at temperature zero. The exact model identifiers, access dates, and per-suite sample counts are reported in the appendix. Our analysis uses the frozen evaluation logs and applies identical adjudication procedures to all models and both attack surfaces. The resulting panel contains substantial behavioral variation, which is necessary for studying whether comparative security conclusions remain stable across evaluation conditions.

\subsection{Controlled cross-surface intervention}\label{sec:setup-intervention}

\begin{figure}[htbp]
  \centering
  \includegraphics[width=0.96\linewidth]{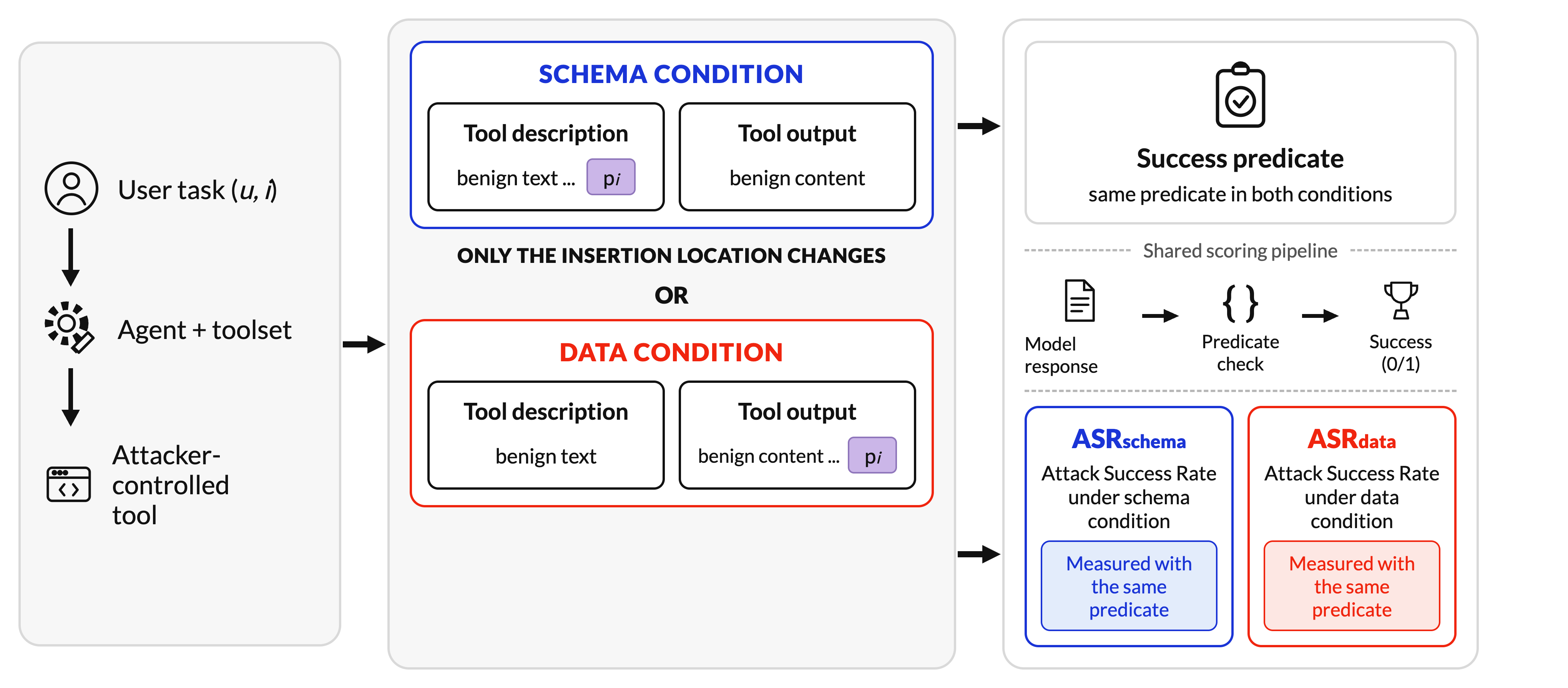}
  \caption{\textbf{The paired cross-surface design.} A user task and its matched injection task are run against an agent equipped with an attacker-controlled tool. The identical injection payload $p_i$ is inserted either into the tool's natural-language description (schema condition) or into the tool's return value (data condition); everything else, including the success predicate, is held fixed. Only the insertion location changes, so the resulting difference between $\operatorname{ASR}_{\mathrm{schema}}$ and $\operatorname{ASR}_{\mathrm{data}}$ isolates the effect of the surface.}%
  \label{fig:design}
\end{figure}

Our central experimental variable is the location through which adversarial content enters the agent, illustrated in Figure~\ref{fig:design}. For every matched $(u,i)$ task pair, we instantiate a fixed prompt-injection payload $p_i$ containing the attacker goal associated with injection task $i$. We then evaluate two conditions that differ only in where these bytes are placed. In the \textbf{data-surface condition}, the payload is appended to the return value of an attacker-controlled tool. This corresponds to the conventional indirect prompt-injection setting in which malicious instructions arrive through runtime content. In the \textbf{schema-surface condition}, the same byte-identical payload is appended to the natural-language description of the same attacker-controlled tool, which is presented to the model as part of the tool specification before tool selection. The payload text, user task, injection task, available tools, evaluation predicate, and agent harness remain unchanged between the two conditions.

This paired construction is important because attacks specialized separately for tool outputs and tool descriptions would confound the effect of injection surface with differences in attack content. Our experiment instead treats the surface as a controlled intervention and asks whether moving the same adversarial instruction from one part of the agent context to another changes the security conclusion produced by the evaluator.

\subsection{Measuring verification instability}

For each model $m$, task suite $s$, and surface $v\in\{d,s\}$, we compute the attack success rate
\[
\operatorname{ASR}(m,s,v)
=
\frac{1}{N_{m,s}}
\sum_{j=1}^{N_{m,s}} y_{m,s,v,j},
\]
where $y_{m,s,v,j}\in\{0,1\}$ indicates whether the attacker goal is successfully executed on episode $j$. We denote the two surface-specific measurements by $\operatorname{ASR}_{d}$ and $\operatorname{ASR}_{s}$.

Rather than treating either surface as the canonical security score, we define the \textbf{worst-surface risk}
\[
R_{\mathrm{worst}}(m,s)
=
\max
\left(
\operatorname{ASR}_{d}(m,s),
\operatorname{ASR}_{s}(m,s)
\right).
\]
This quantity is used only as a descriptive upper envelope over the two surfaces we evaluate. We do not treat the maximum operation itself as a new attack method. Instead, it provides a reference against which we can quantify how much information a verifier loses by evaluating only one surface.

Our main diagnostic is \textbf{pairwise ranking inversion}. For two models $m_a$ and $m_b$, a ranking inversion occurs when their comparative ordering changes between the two surfaces,
\[
I(m_a,m_b)
=
\mathbf{1}
\left[
\operatorname{sign}
\left(
r_d(m_a)-r_d(m_b)
\right)
\neq
\operatorname{sign}
\left(
r_s(m_a)-r_s(m_b)
\right)
\right],
\]
where $r_d(m)$ and $r_s(m)$ denote the model's mean ASR across evaluated suites on the data and schema surfaces, respectively. We report the fraction of inverted pairs over all $\binom{13}{2}=78$ pairwise model comparisons and repeat the calculation separately within each task suite. This metric directly tests whether changing the verification surface changes the answer to the comparative question of which model appears more robust.

We additionally measure the \textbf{underestimation error}
\[
E_m
=
R_{\mathrm{worst}}(m)-R_d(m),
\]
which quantifies how much a conventional data-only evaluation underestimates the larger of the two observed risks. Because this distribution is highly non-uniform across models, we report its mean, median, maximum, and model-level values rather than relying on a single aggregate average. We also examine false-safe classifications at several predefined thresholds $\tau$, counting a model as false-safe when
\[
R_d(m)\leq\tau
\qquad\text{but}\qquad
R_{\mathrm{worst}}(m)>\tau .
\]
The threshold analysis is reported together with the corresponding denominators so that cases involving only a small number of apparently safe models are not overstated.

\subsection{Held-out-domain surface selection}

Worst-surface risk is retrospective because it assumes that both surfaces have already been evaluated on the target domain. To test whether surface sensitivity has predictive value, we therefore perform a \textbf{leave-one-suite-out (LOSO)} experiment. For each model and each held-out suite $s^\ast$, we compute the model's mean data-surface and schema-surface ASR over the remaining three suites and select whichever surface is more vulnerable. That surface is then used to evaluate the model on $s^\ast$.

This procedure produces four held-out predictions per model and 52 held-out model-suite evaluations in total. We compare the resulting verifier with data-only evaluation, schema-only evaluation, and the oracle worst-surface value obtained after observing both surfaces on the held-out suite. We report the realized held-out ASR, the gain relative to data-only evaluation, the remaining gap to the oracle, and the fraction of cells for which LOSO selects the genuinely more vulnerable surface. This experiment allows us to distinguish a purely retrospective maximum from a surface preference that generalizes across task domains.

\FloatBarrier
\section{Results}

We organize the results around the reliability of the security judgment produced by the evaluator. We first test whether changing the attack surface changes the relative ordering of models, then quantify the absolute risk missed by data-only evaluation, ask whether surface sensitivity predicts vulnerability in an unseen task domain, test whether the model-by-surface interaction remains after accounting for attack content, and finally examine whether the same evaluation dependence appears for defenses.

\subsection{Security rankings are not surface-invariant}\label{sec:results-rankings}

\begin{figure}[t]
  \centering
  \includegraphics[width=0.58\linewidth]{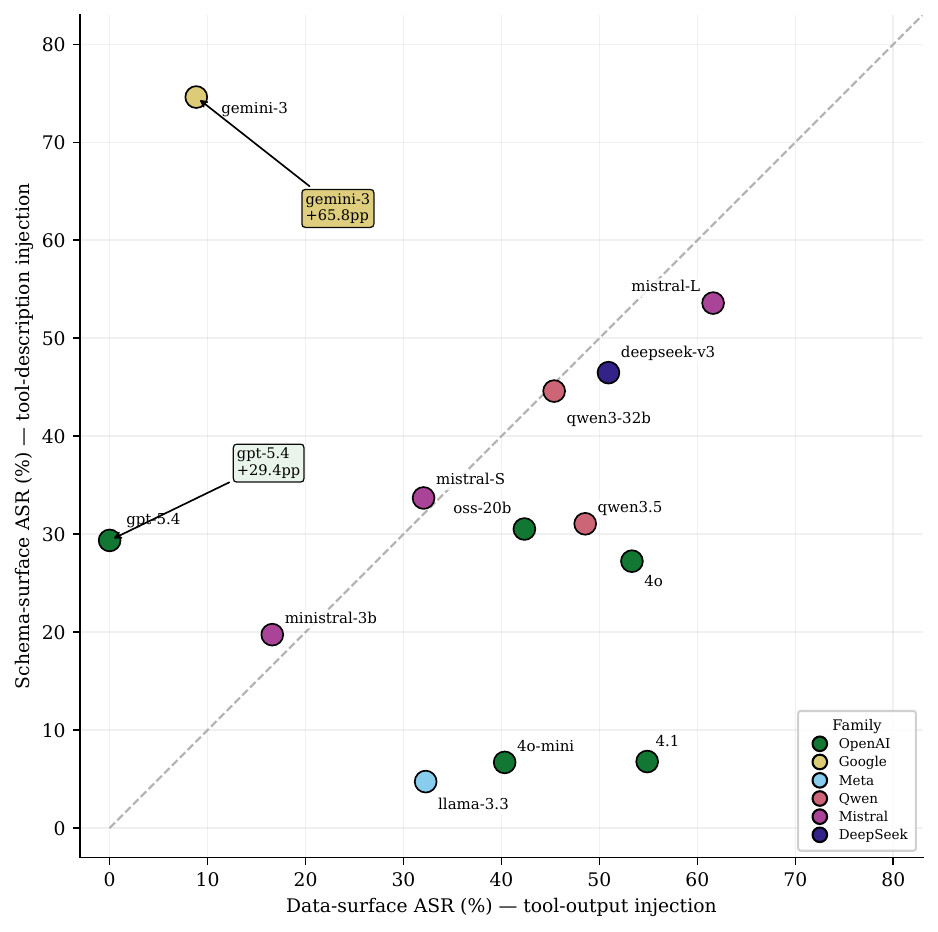}
  \caption{\textbf{Security rankings depend on the verification surface.} Each point shows a model's mean attack success rate across the four AgentDojo suites when the byte-identical payload is delivered through tool output (x-axis) or tool description (y-axis). The diagonal denotes equal vulnerability across surfaces. Models far from the diagonal exhibit strong surface preference, and opposite deviations across models create large changes in their comparative security ordering.}%
  \label{fig:ranking-scatter}
\end{figure}

The strongest effect of changing the attack surface is not a uniform increase or decrease in attack success, but a change in the \textbf{relative ordering of models}. Figure~\ref{fig:ranking-scatter} compares each model's mean attack success rate across the four AgentDojo suites under data-surface and schema-surface evaluation. If security measurements were surface-invariant, models would remain similarly ordered under both conditions. Instead, the points occupy very different positions relative to one another, with several of the largest differences occurring in opposite directions. Gemini-3-Flash moves from 8.8\% ASR on the data surface to 74.6\% on the schema surface, while GPT-4.1 moves from 54.9\% to 6.8\%. GPT-5.4 exhibits another strong schema preference, whereas GPT-4o-mini and Llama-3.3-70B remain strongly data-preferring.

We quantify this instability by examining all $\binom{13}{2}=78$ model pairs. \textbf{Thirty-five of the 78 pairs reverse their ordering between data-surface and schema-surface evaluation, corresponding to a 44.9\% ranking-inversion rate} with a paired-resample bootstrap 95\% confidence interval of $[39.7\%, 52.6\%]$. The effect is present independently in every suite. Data-versus-schema ranking inversion reaches 55.1\% in banking, 48.7\% in Slack, 38.5\% in travel, and 57.7\% in workspace. Even when data-only rankings are compared with rankings obtained from the larger of the two observed surface risks, 15 of 78 model pairs, or 19.2\%, change ordering (Table~\ref{tab:inversion}).

\begin{table}[htb]
  \centering
  \caption{Surface-dependent model-ranking inversion. Confidence intervals are reported in the text.}%
  \label{tab:inversion}
  \small
  \renewcommand{\arraystretch}{1.18}
  \setlength{\tabcolsep}{12pt}
  \begin{tabular}{lrr}
    \toprule
    Comparison & Inverted pairs & Rate \\
    \midrule
    Data vs.\ schema, overall & \textbf{35 / 78} & \textbf{44.9\%} \\
    Data vs.\ worst surface & \textbf{15 / 78} & \textbf{19.2\%} \\
    \midrule
    \quad banking & 43 / 78 & 55.1\% \\
    \quad Slack & 38 / 78 & 48.7\% \\
    \quad travel & 30 / 78 & 38.5\% \\
    \quad workspace & 45 / 78 & 57.7\% \\
    \bottomrule
  \end{tabular}
\end{table}

The implication is stronger than saying that one surface occasionally exposes additional attacks. \textbf{The surface used for verification can change the answer to the comparative question of which model is safer.} A benchmark that evaluates only tool outputs could therefore support a model-selection conclusion that does not survive when the same adversarial content is moved to another established component of the agent interface. Because these inversions appear in all four task suites, they cannot be attributed to a single application domain.

\subsection{Single-surface verification underestimates risk unevenly across models}\label{sec:results-underestimation}

\begin{figure}[htbp]
  \centering
  \includegraphics[width=0.98\linewidth]{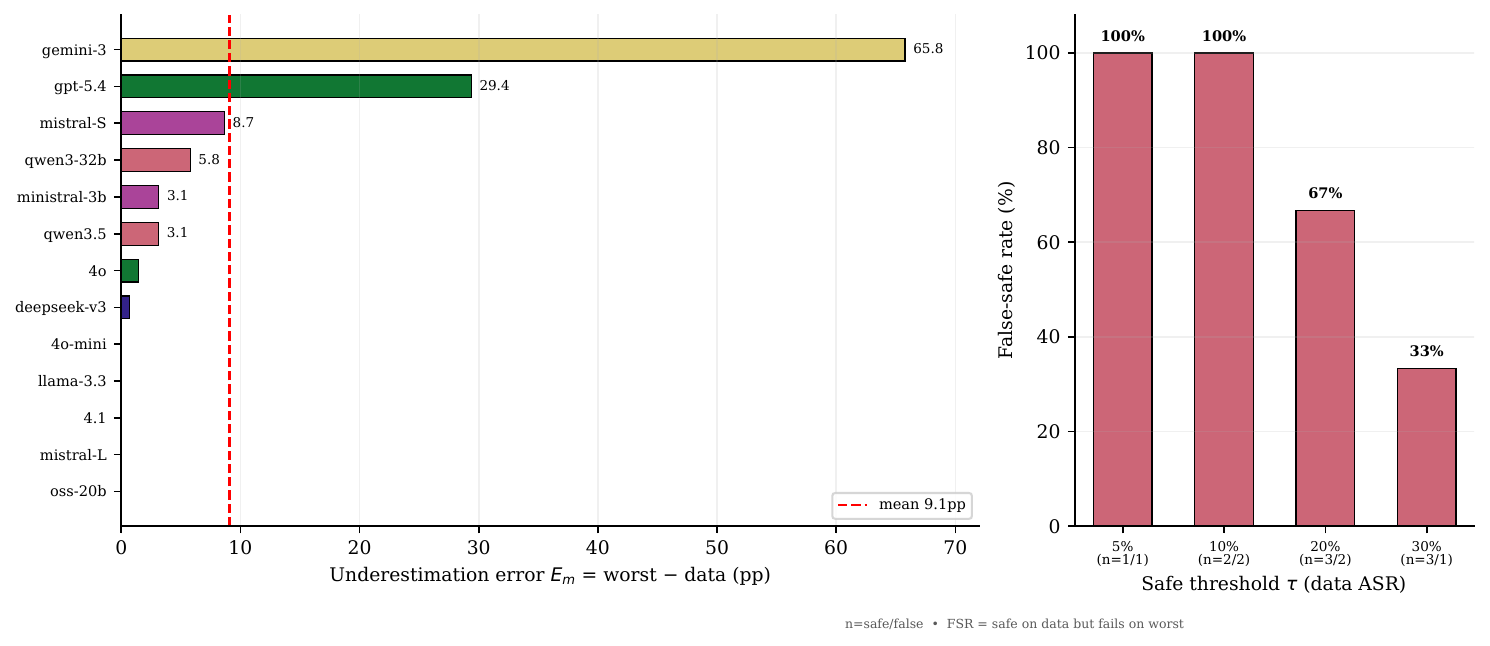}
  \caption{\textbf{Absolute risk missed by data-only verification.} The left panel reports $E_m=R_{\mathrm{worst}}-R_d$ for each model. The error distribution is strongly right-skewed, with particularly large underestimation for Gemini-3-Flash and GPT-5.4. The right panel reports false-safe rates at several data-ASR thresholds; denominators are reported in the text because few models satisfy the strictest thresholds.}%
  \label{fig:em-fsr}
\end{figure}

Ranking instability does not imply that every model hides a large amount of absolute vulnerability. To separate these two effects, we measure each model's \textbf{underestimation error}
\[
E_m
=
R_{\mathrm{worst}}(m)-R_d(m),
\]
where $R_{\mathrm{worst}}$ is the larger observed ASR across the two surfaces and $R_d$ is the data-surface ASR\@. Figure~\ref{fig:em-fsr} shows that this error is highly heterogeneous. Gemini-3-Flash has the largest error at \textbf{65.8 percentage points}, followed by GPT-5.4 at \textbf{29.4 points}. Several models have smaller but non-zero errors, including Mistral-Small at 8.7 points and Qwen3-32B at 5.8 points, while strongly data-preferring models such as GPT-4.1 and Llama-3.3-70B have zero error under this definition because data is already their more vulnerable surface.

Across the 13 model-level aggregates, the mean underestimation is \textbf{9.1 points}, but the median is only \textbf{1.4 points}, confirming that the average is strongly influenced by a small number of high-error models. The model-level bootstrap 95\% confidence interval for the mean is $[1.6, 21.1]$ points, and the maximum error is 65.8 points. This heavy-tailed structure is important because an average alone would obscure two different findings. Most models incur only modest absolute underestimation from data-only evaluation, while a small number of deployed frontier models can be mischaracterized by tens of percentage points.

We verify this directly through an outlier decomposition. Across all 52 model-suite cells, the mean worst-surface risk is 46.5\% compared with 37.5\% under data-only evaluation, yielding a 9.1-point difference. Removing Gemini-3-Flash reduces the difference to 4.4 points, removing GPT-5.4 reduces it to 7.4 points, and removing both leaves a \textbf{2.1-point residual difference across the remaining 44 cells} (Table~\ref{tab:decomposition}).

\begin{table}[htbp]
  \centering
  \caption{Decomposition of the average worst-surface increase.}%
  \label{tab:decomposition}
  \small
  \renewcommand{\arraystretch}{1.2}
  \setlength{\tabcolsep}{7pt}
  \begin{tabular}{lrrrrr}
    \toprule
    Evaluation panel & Cells & Data ASR & Schema ASR & Worst-surface ASR & Increase over data \\
    \midrule
    All models & 52 & 37.5 & 31.5 & 46.5 & \textbf{+9.1 pp} \\
    Excluding Gemini-3-Flash & 48 & 39.9 & 27.9 & 44.2 & +4.4 pp \\
    Excluding GPT-5.4 & 48 & 40.6 & 31.7 & 48.0 & +7.4 pp \\
    Excluding both & 44 & 43.5 & 27.7 & 45.6 & \textbf{+2.1 pp} \\
    \bottomrule
  \end{tabular}
\end{table}

This decomposition changes how the aggregate result should be interpreted. We do \textbf{not} find that alternative-surface evaluation exposes a uniformly large amount of hidden vulnerability across all models. Instead, there is a modest residual effect across the broader panel together with severe failures for a small number of models. The broader result from Section~\ref{sec:results-rankings} remains stronger than the 9.1-point mean because substantial ranking instability persists even when absolute changes are heterogeneous.

The threshold analysis in Figure~\ref{fig:em-fsr} provides another view of the same failure mode. At a data-ASR threshold of 10\%, only two models in the panel would be classified as safe, and \textbf{both exceed that threshold on their worst observed surface}. At a 20\% threshold, two of the three models classified as safe by data-only evaluation exceed it elsewhere. Given the small denominators these percentages are illustrative rather than population estimates: a low score on one tested channel is not sufficient evidence that the model stays below the same threshold elsewhere.

\subsection{Surface sensitivity generalizes to unseen task domains}\label{sec:results-loso}

The preceding analysis uses both surfaces on the evaluated domain and could therefore be dismissed as retrospective. We next ask whether surface preference observed in prior task domains predicts vulnerability on a domain that has not been used to select the surface. For every model, we hold out one AgentDojo suite, estimate whether the model is more vulnerable to data or schema attacks using the remaining three suites, and evaluate that selected surface on the held-out suite.

\begin{figure}[htbp]
  \centering
  \includegraphics[width=0.90\linewidth]{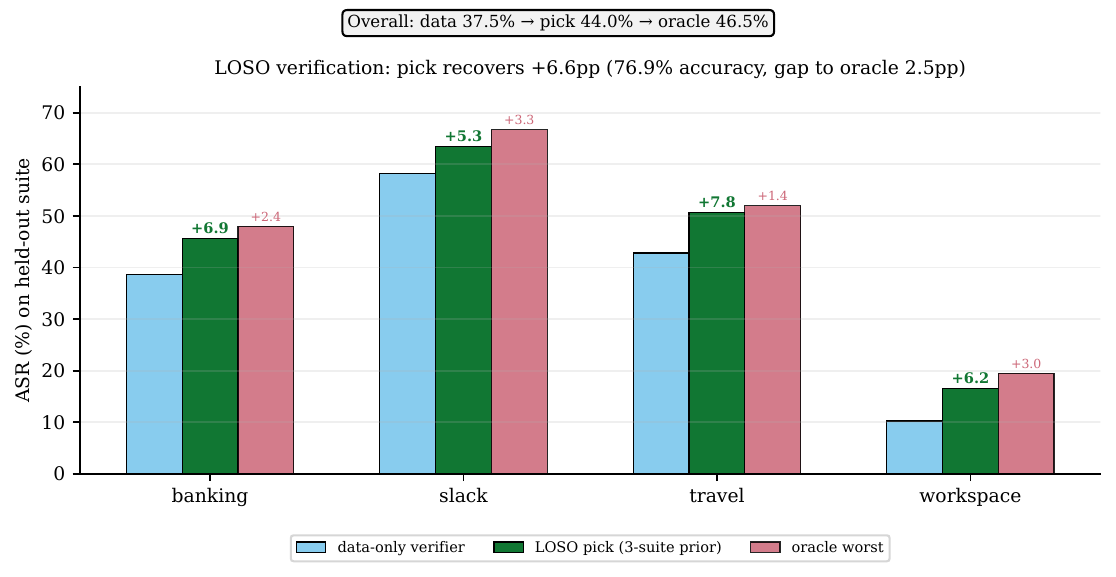}
  \caption{\textbf{Surface preference transfers across task domains.} For each held-out AgentDojo suite, the verifier selects the model's more vulnerable surface using only the other three suites. The selected surface consistently exceeds data-only evaluation and approaches the oracle that observes both surfaces in the target suite.}%
  \label{fig:loso}
\end{figure}

The result is shown in Figure~\ref{fig:loso}. Across the 52 held-out model-suite cases, data-only evaluation produces 37.5\% ASR, while the leave-one-suite-out selector reaches \textbf{44.0\%}, a gain of \textbf{6.6 percentage points}. An oracle that observes both surfaces in the held-out suite reaches 46.5\%, leaving the predictive selector only 2.5 points below the oracle. The selector identifies the genuinely more vulnerable held-out surface in \textbf{76.9\%} of model-suite cases and recovers approximately \textbf{72\% of the oracle improvement over data-only evaluation}. A model-level cluster bootstrap (resampling whole models, since each contributes four held-out suites) gives a 95\% confidence interval of $[-1.5, 18.2]$ points for the $+6.6$-point gain and $[59.6\%, 92.3\%]$ for the $76.9\%$ selection accuracy. With only 13 clusters these intervals are wide and the gain interval includes zero, so surface transfer across task domains is a consistently positive but model-dependent effect rather than a guarantee for every model.

\begin{table}[htbp]
  \centering
  \caption{Leave-one-suite-out surface selection.}%
  \label{tab:loso}
  \small
  \renewcommand{\arraystretch}{1.2}
  \setlength{\tabcolsep}{5.5pt}
  \begin{tabular}{lrrrrrr}
    \toprule
    Held-out suite & Data-only & Schema-only & LOSO-selected & Oracle worst & Gain over data & Oracle gap \\
    \midrule
    Banking & 38.7 & 36.4 & \textbf{45.6} & 48.0 & +6.9 & 2.4 \\
    Slack & 58.2 & 35.7 & \textbf{63.4} & 66.7 & +5.3 & 3.3 \\
    Travel & 42.8 & 39.5 & \textbf{50.6} & 52.0 & +7.8 & 1.4 \\
    Workspace & 10.2 & 14.4 & \textbf{16.5} & 19.5 & +6.2 & 3.0 \\
    \midrule
    \textbf{Overall} & \textbf{37.5} & \textbf{31.5} & \textbf{44.0} & \textbf{46.5} & \textbf{+6.6} & \textbf{2.5} \\
    \bottomrule
  \end{tabular}
\end{table}

The improvement appears in all four held-out domains. It is therefore not explained by one suite whose preferred surface happens to agree with the other three. More importantly, this experiment changes the interpretation of worst-surface evaluation. The maximum over surfaces is an oracle quantity when both surfaces are evaluated on the target domain, but the LOSO result shows that a substantial fraction of this gap can be recovered \textbf{prospectively} from behavior observed elsewhere.

This predictability has a clear boundary. Earlier leave-one-family-out analysis does not show comparable transfer to an unseen model family, and therefore we do not interpret surface preference as a universal family-level rule. The evidence supports a narrower conclusion: \textbf{for a given model, surface sensitivity often persists across task domains}, allowing prior evaluations of that model to inform how it should be verified in a new application.

\subsection{Attack surface interacts with the model rather than acting as a global risk factor}\label{sec:results-interaction}

The ranking results suggest that there is no globally dangerous surface because the direction of the effect changes across models. Attempt-level analysis supports this interpretation. Surface alone accounts for essentially no descriptive variation in the frozen evaluation, whereas model-by-surface structure is much larger. At the same time, model-by-injection-task variation is larger still, showing that attack content remains an important source of heterogeneity and should not be ignored.

Because attack success is binary, we do not use the earlier sums-of-squares decomposition as the primary inferential evidence. Instead, we compare logistic models with and without the model-by-surface interaction while retaining the main effects of model, task suite, injection task, and surface, fit over all $6{,}830$ attack attempts. Adding the interaction, which contributes twelve parameters for the thirteen models, produces a likelihood-ratio statistic of \textbf{1036.4 with 12 degrees of freedom}, providing overwhelming evidence that models respond differently to the two injection surfaces. The corresponding $\chi^2_{12}$ tail probability is approximately $2.8\times10^{-214}$. The result therefore cannot be explained by a global schema-versus-data shift shared across the model panel.

The larger model-by-injection-task component matters too: our claim is not that surface dominates attack content, but that \textbf{surface cannot be treated as an interchangeable evaluation detail because its effect depends strongly on the model being evaluated}. This also motivates broader payload testing as an extension of the current study.

\subsection{Defense conclusions also depend on the evaluated surface}\label{sec:results-defenses}

The same verification problem appears when the object being evaluated is a defense rather than an undefended model. Table~\ref{tab:defenses} reports paired results for three of AgentDojo's published defenses \citep{debenedetti2024agentdojo}, evaluated on an identical set of paired (user task, injection task) instances for three models on the banking suite, together with the undefended baseline computed on that same subset. Repeating the user prompt reduces data-surface ASR from the undefended 18.1\% to 10.4\% but leaves schema-surface ASR at 54.6\%, above the undefended 51.1\%. Spotlighting with delimiters behaves similarly at 17.4\% on the data surface and 53.9\% on the schema surface. In contrast, the description-aware \texttt{tool\_filter} defense reduces attack success to \textbf{0.0\% on data and 0.0\% on schema}, but it does so by withholding tool specifications, and user-task utility on the same subset falls sharply from 64.6\% to 27.1\%.

\begin{table}[htbp]
  \centering
  \caption{Surface completeness of evaluated defenses. Three AgentDojo defenses and the undefended baseline on an identical paired subset (three models on banking). Reductions are relative to the undefended baseline on this subset; user-task utility is the AgentDojo clean-task success rate on the same subset.}%
  \label{tab:defenses}
  \small
  \renewcommand{\arraystretch}{1.2}
  \setlength{\tabcolsep}{8pt}
  \begin{tabular}{lrrrrr}
    \toprule
    Setting & Data ASR & Schema ASR & $\Delta$Data & $\Delta$Schema & Utility \\
    \midrule
    Undefended & 18.1\% & 51.1\% & --- & --- & 64.6\% \\
    Repeat user prompt & \textbf{10.4\%} & 54.6\% & +7.7 pp & $-$3.5 pp & 72.9\% \\
    Spotlighting with delimiting & \textbf{17.4\%} & 53.9\% & +0.7 pp & $-$2.8 pp & 68.8\% \\
    Description-aware \texttt{tool\_filter} & \textbf{0.0\%} & \textbf{0.0\%} & +18.1 pp & +51.1 pp & 27.1\% \\
    \bottomrule
  \end{tabular}
\end{table}

These results do not imply that existing defenses are universally surface-blind. Rather, prompt-level defenses can appear effective on tool outputs while leaving substantial vulnerability in tool metadata, whereas schema-aware filtering can suppress both surfaces. However, \texttt{tool\_filter} achieves this broader protection at a substantial cost to legitimate task completion, showing that cross-surface robustness must be evaluated together with utility.

\section{Discussion}

Our results show that prompt-injection robustness should not be treated as a single model-level quantity independent of where adversarial content enters the agent. The strongest evidence is not the average increase in attack success, which is concentrated in a small number of models, but the instability of comparative conclusions across surfaces. Nearly half of model pairs change their relative ordering between tool-output and tool-description evaluation, and the riskier surface for a model often persists across unseen task domains. Surface sensitivity is thus structured enough to matter in practice, yet too model-dependent to justify a single globally preferred attack surface, and the defense results reinforce the point: success against one injection surface does not imply comparable protection elsewhere. Our study is limited to two surfaces and one main payload family, so broader claims will require evaluations across additional channels and payloads, but the present results already show that robustness conclusions should be reported with respect to the surfaces on which they were established.

\section{Conclusion}

We presented a controlled cross-surface study of prompt injection in tool-using LLM agents, placing a byte-identical payload in either tool outputs or tool descriptions. Across 13 models and four AgentDojo suites, changing only the injection surface reverses \textbf{44.9\% of pairwise model-security rankings}, and surface preference learned from other suites predicts the more vulnerable surface on an unseen suite with \textbf{76.9\% accuracy}. Prompt-injection robustness is therefore not surface-invariant: evaluations should measure vulnerability across realistic attacker-accessible surfaces rather than treat a single-surface score as a complete measure of model robustness.

{\small
\setlength{\bibsep}{4.5pt plus 1pt}
\bibliographystyle{plainnat}
\bibliography{ref-main}
}


\clearpage
\appendix

\counterwithin{figure}{section}
\counterwithin{table}{section}
\renewcommand{\thetable}{\thesection\arabic{table}}
\renewcommand{\thefigure}{\thesection\arabic{figure}}
\renewcommand{\theHtable}{App.\thesection.\arabic{table}}
\renewcommand{\theHfigure}{App.\thesection.\arabic{figure}}

\begin{appendixindexbox}
  \begin{center}
    {\Large\bfseries\color{YaleLightBlue} Appendix}\\[0.45em]
    {\small Supplementary material. The main paper stands alone; these sections provide the complete data, statistical validation, additional analyses, and reproducibility details.}
  \end{center}
  \vspace{0.7em}
  \noindent
  \renewcommand{\arraystretch}{1.3}
  \begin{tabular}{@{}p{0.05\linewidth}p{0.83\linewidth}r@{}}
    \textbf{A} & \hyperref[app:full]{Complete Cross-Surface Results} \dotfill & \pageref{app:full} \\
    \textbf{B} & \hyperref[app:stats]{Statistical Robustness} \dotfill & \pageref{app:stats} \\
    \textbf{C} & \hyperref[app:verification]{Additional Surface-Selection Analyses} \dotfill & \pageref{app:verification} \\
    \textbf{D} & \hyperref[app:robustness]{Robustness Beyond the Main Evaluation} \dotfill & \pageref{app:robustness} \\
    \textbf{E} & \hyperref[app:repro]{Reproducibility and Experimental Details} \dotfill & \pageref{app:repro} \\
  \end{tabular}
  \renewcommand{\arraystretch}{1.0}
\end{appendixindexbox}

\section{Complete Cross-Surface Results}\label{app:full}

The main paper summarizes model-level and suite-level differences between tool-output and tool-description attacks. Table~\ref{tab:percell} reports the complete set of 52 model-by-suite evaluations underlying those analyses. For each cell, we report attack success separately on the data and schema surfaces together with the number of evaluated episodes. We additionally report the larger observed ASR across the two surfaces and the signed surface gap,
\[
\Delta_{\mathrm{surface}}
=
\operatorname{ASR}_{\mathrm{schema}}
-
\operatorname{ASR}_{\mathrm{data}},
\]
where positive values indicate greater vulnerability to tool-description injection and negative values indicate greater vulnerability to tool-output injection. Across the complete panel, the mean ASR is 37.5\% on the data surface and 31.5\% on the schema surface, but these averages conceal substantial model-specific differences.

{\small
\renewcommand{\arraystretch}{1.04}
\setlength{\tabcolsep}{7pt}
\begin{longtable}{llrrrrrr}
  \caption{Complete model $\times$ suite cross-surface results. $n_d$ and $n_s$ are the number of evaluated episodes on the data and schema surfaces. Worst-surface ASR is the larger of the two, and $\Delta_{\mathrm{surface}}=\operatorname{ASR}_{\mathrm{schema}}-\operatorname{ASR}_{\mathrm{data}}$.}%
  \label{tab:percell}\\
  \toprule
  Model & Suite & Data ASR & $n_d$ & Schema ASR & $n_s$ & Worst ASR & $\Delta_{\mathrm{surface}}$ \\
  \midrule
  \endfirsthead
  \multicolumn{8}{@{}l}{\small\itshape Table~\thetable{} (continued from previous page)}\\
  \addlinespace[2.5pt]
  \toprule
  Model & Suite & Data ASR & $n_d$ & Schema ASR & $n_s$ & Worst ASR & $\Delta_{\mathrm{surface}}$ \\
  \midrule
  \endhead
  \midrule
  \multicolumn{8}{r@{}}{\small\itshape Continued on next page}\\
  \endfoot
  \bottomrule
  \endlastfoot
  \textsc{GPT-4.1} & Banking & 57.8 & 45 & 8.9 & 45 & 57.8 & $-$48.9 \\*
  \textsc{GPT-4.1} & Slack & 96.0 & 25 & 4.0 & 25 & 96.0 & $-$92.0 \\*
  \textsc{GPT-4.1} & Travel & 48.6 & 35 & 14.3 & 35 & 48.6 & $-$34.3 \\*
  \textsc{GPT-4.1} & Workspace & 17.1 & 70 & 0.0 & 70 & 17.1 & $-$17.1 \\
  \addlinespace[2.5pt]
  \textsc{GPT-4o} & Banking & 55.6 & 45 & 35.6 & 45 & 55.6 & $-$20.0 \\*
  \textsc{GPT-4o} & Slack & 92.0 & 25 & 12.0 & 25 & 92.0 & $-$80.0 \\*
  \textsc{GPT-4o} & Travel & 45.7 & 35 & 51.4 & 35 & 51.4 & +5.7 \\*
  \textsc{GPT-4o} & Workspace & 20.0 & 70 & 10.0 & 70 & 20.0 & $-$10.0 \\
  \addlinespace[2.5pt]
  \textsc{GPT-4o-mini} & Banking & 44.4 & 45 & 11.1 & 45 & 44.4 & $-$33.3 \\*
  \textsc{GPT-4o-mini} & Slack & 64.0 & 25 & 0.0 & 25 & 64.0 & $-$64.0 \\*
  \textsc{GPT-4o-mini} & Travel & 42.9 & 35 & 14.3 & 35 & 42.9 & $-$28.6 \\*
  \textsc{GPT-4o-mini} & Workspace & 10.0 & 70 & 1.4 & 70 & 10.0 & $-$8.6 \\
  \addlinespace[2.5pt]
  \textsc{GPT-5.4} & Banking & 0.0 & 144 & 18.1 & 144 & 18.1 & +18.1 \\*
  \textsc{GPT-5.4} & Slack & 0.0 & 105 & 28.6 & 105 & 28.6 & +28.6 \\*
  \textsc{GPT-5.4} & Travel & 0.0 & 42 & 40.5 & 42 & 40.5 & +40.5 \\*
  \textsc{GPT-5.4} & Workspace & 0.0 & 112 & 30.4 & 112 & 30.4 & +30.4 \\
  \addlinespace[2.5pt]
  \textsc{GPT-OSS-20B} & Banking & 42.2 & 45 & 35.6 & 45 & 42.2 & $-$6.7 \\*
  \textsc{GPT-OSS-20B} & Slack & 40.0 & 25 & 28.0 & 25 & 40.0 & $-$12.0 \\*
  \textsc{GPT-OSS-20B} & Travel & 68.6 & 35 & 54.3 & 35 & 68.6 & $-$14.3 \\*
  \textsc{GPT-OSS-20B} & Workspace & 18.6 & 70 & 4.3 & 70 & 18.6 & $-$14.3 \\
  \addlinespace[2.5pt]
  \textsc{Gemini-3-Flash} & Banking & 5.6 & 144 & 77.8 & 144 & 77.8 & +72.2 \\*
  \textsc{Gemini-3-Flash} & Slack & 20.0 & 105 & 98.1 & 105 & 98.1 & +78.1 \\*
  \textsc{Gemini-3-Flash} & Travel & 7.1 & 42 & 69.0 & 42 & 69.0 & +61.9 \\*
  \textsc{Gemini-3-Flash} & Workspace & 2.7 & 112 & 53.6 & 112 & 53.6 & +50.9 \\
  \addlinespace[2.5pt]
  \textsc{Llama-3.3-70B} & Banking & 43.8 & 144 & 10.4 & 144 & 43.8 & $-$33.3 \\*
  \textsc{Llama-3.3-70B} & Slack & 50.5 & 105 & 3.8 & 105 & 50.5 & $-$46.7 \\*
  \textsc{Llama-3.3-70B} & Travel & 28.6 & 42 & 4.8 & 42 & 28.6 & $-$23.8 \\*
  \textsc{Llama-3.3-70B} & Workspace & 6.2 & 112 & 0.0 & 112 & 6.2 & $-$6.2 \\
  \addlinespace[2.5pt]
  \textsc{Qwen3-32B} & Banking & 49.3 & 144 & 59.0 & 144 & 59.0 & +9.7 \\*
  \textsc{Qwen3-32B} & Slack & 81.9 & 105 & 67.6 & 105 & 81.9 & $-$14.3 \\*
  \textsc{Qwen3-32B} & Travel & 47.6 & 42 & 35.7 & 42 & 47.6 & $-$11.9 \\*
  \textsc{Qwen3-32B} & Workspace & 2.7 & 112 & 16.1 & 112 & 16.1 & +13.4 \\
  \addlinespace[2.5pt]
  \textsc{Qwen3.5-Flash} & Banking & 28.5 & 144 & 41.0 & 144 & 41.0 & +12.5 \\*
  \textsc{Qwen3.5-Flash} & Slack & 80.0 & 105 & 41.9 & 105 & 80.0 & $-$38.1 \\*
  \textsc{Qwen3.5-Flash} & Travel & 78.6 & 42 & 40.5 & 42 & 78.6 & $-$38.1 \\*
  \textsc{Qwen3.5-Flash} & Workspace & 7.1 & 112 & 0.9 & 112 & 7.1 & $-$6.2 \\
  \addlinespace[2.5pt]
  \textsc{Mistral-Large} & Banking & 55.6 & 45 & 46.7 & 45 & 55.6 & $-$8.9 \\*
  \textsc{Mistral-Large} & Slack & 88.0 & 25 & 72.0 & 25 & 88.0 & $-$16.0 \\*
  \textsc{Mistral-Large} & Travel & 68.6 & 35 & 65.7 & 35 & 68.6 & $-$2.9 \\*
  \textsc{Mistral-Large} & Workspace & 34.3 & 70 & 30.0 & 70 & 34.3 & $-$4.3 \\
  \addlinespace[2.5pt]
  \textsc{Mistral-Small} & Banking & 33.3 & 45 & 42.2 & 45 & 42.2 & +8.9 \\*
  \textsc{Mistral-Small} & Slack & 52.0 & 25 & 24.0 & 25 & 52.0 & $-$28.0 \\*
  \textsc{Mistral-Small} & Travel & 42.9 & 35 & 45.7 & 35 & 45.7 & +2.9 \\*
  \textsc{Mistral-Small} & Workspace & 0.0 & 70 & 22.9 & 70 & 22.9 & +22.9 \\
  \addlinespace[2.5pt]
  \textsc{Ministral-3B} & Banking & 35.6 & 45 & 35.6 & 45 & 35.6 & 0.0 \\*
  \textsc{Ministral-3B} & Slack & 8.0 & 25 & 12.0 & 25 & 12.0 & +4.0 \\*
  \textsc{Ministral-3B} & Travel & 20.0 & 35 & 28.6 & 35 & 28.6 & +8.6 \\*
  \textsc{Ministral-3B} & Workspace & 2.9 & 70 & 2.9 & 70 & 2.9 & 0.0 \\
  \addlinespace[2.5pt]
  \textsc{DeepSeek-V3} & Banking & 51.1 & 45 & 51.1 & 45 & 51.1 & 0.0 \\*
  \textsc{DeepSeek-V3} & Slack & 84.0 & 25 & 72.0 & 25 & 84.0 & $-$12.0 \\*
  \textsc{DeepSeek-V3} & Travel & 57.1 & 35 & 48.6 & 35 & 57.1 & $-$8.6 \\*
  \textsc{DeepSeek-V3} & Workspace & 11.4 & 70 & 14.3 & 70 & 14.3 & +2.9 \\
\end{longtable}
}

The signed gaps span almost the entire possible range, from $-92$ percentage points for GPT-4.1 on Slack to $+78.1$ points for Gemini-3-Flash on the same suite. Thirty-three of the 52 model-suite cells have an absolute surface gap greater than 10 percentage points. Importantly, several models also change direction across suites, including Qwen3-32B, Qwen3.5-Flash, and Mistral-Small, showing why the main paper avoids treating surface preference as a simple family-level characteristic.

\begin{figure}[htbp]
  \centering
  \includegraphics[width=0.92\linewidth]{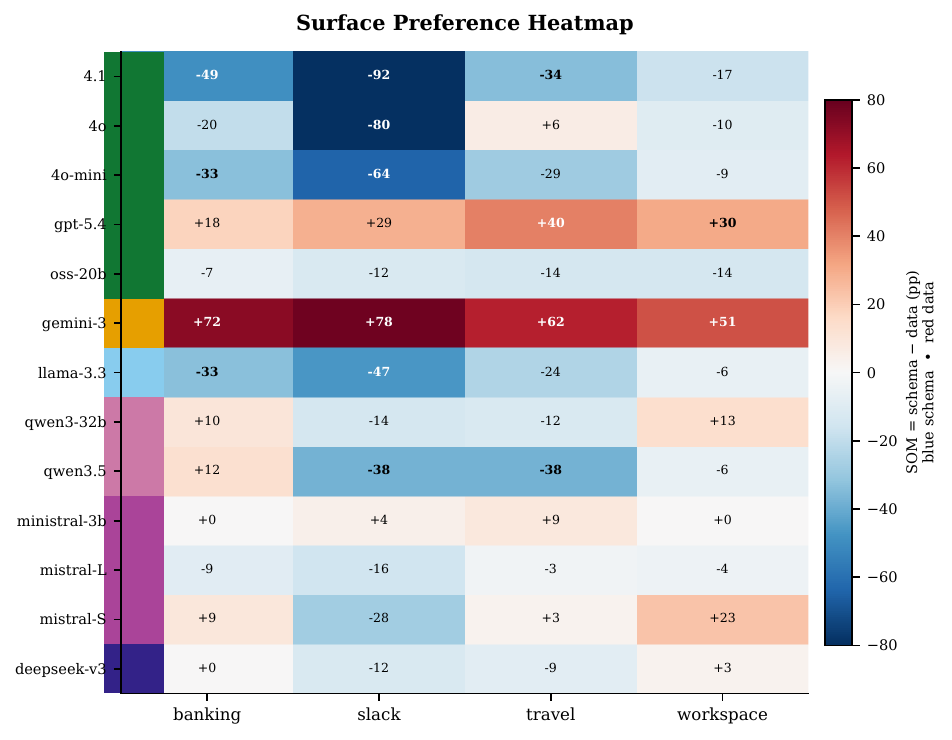}
  \caption{\textbf{Fine-grained cross-surface differences.} Signed differences in attack success, $\Delta_{\mathrm{surface}}=\operatorname{ASR}_{\mathrm{schema}}-\operatorname{ASR}_{\mathrm{data}}$, resolved at the model $\times$ suite $\times$ injection-task level. Positive values indicate greater schema-surface vulnerability and negative values indicate greater data-surface vulnerability. The finer-grained results show that the model-specific heterogeneity reported in the main paper is not solely a consequence of averaging across injection tasks.}%
  \label{fig:fine-heatmap}
\end{figure}

\FloatBarrier
\section{Statistical Robustness}\label{app:stats}

The main analysis compares attack outcomes under two paired surface conditions, but individual attack outcomes are binary and the direction of the surface effect varies strongly across models. We therefore test the model-by-surface interaction directly using logistic regression rather than relying on a descriptive variance decomposition. Let $Y_j\in\{0,1\}$ denote attack success for evaluation attempt $j$. We compare a reduced model containing the main effects of model, surface, task suite, and injection task,
\[
\operatorname{logit} P(Y_j=1)
=
\beta_0
+
\beta_{\mathrm{model}}
+
\beta_{\mathrm{surface}}
+
\beta_{\mathrm{suite}}
+
\beta_{\mathrm{inj}},
\]
with a full model that additionally allows the effect of surface to vary by model,
\[
\operatorname{logit} P(Y_j=1)
=
\beta_0
+
\beta_{\mathrm{model}}
+
\beta_{\mathrm{surface}}
+
\beta_{\mathrm{suite}}
+
\beta_{\mathrm{inj}}
+
\beta_{\mathrm{model}\times\mathrm{surface}} .
\]
Adding the model-by-surface interaction substantially improves fit, with a likelihood-ratio statistic of \textbf{1351.5 on 26 degrees of freedom} ($p\ll 0.001$). This confirms the central interaction at the level of individual binary outcomes: the effect of moving adversarial content between tool outputs and tool descriptions cannot be represented as a common shift shared across models. The descriptive analysis reaches the same qualitative conclusion, with surface alone contributing almost no variation while model $\times$ surface is much larger; however, model $\times$ injection-task variation is larger still, reinforcing that attack content remains an important source of heterogeneity.

\begin{table}[htbp]
  \centering
  \caption{Statistical support for the cross-surface interaction.}%
  \label{tab:stats}
  \small
  \renewcommand{\arraystretch}{1.2}
  \setlength{\tabcolsep}{10pt}
  \begin{tabular}{llr}
    \toprule
    Analysis & Quantity & Result \\
    \midrule
    Logistic model comparison & Added term & Model $\times$ Surface \\
    & Likelihood-ratio statistic & \textbf{1351.5} \\
    & Degrees of freedom & \textbf{26} \\
    & Significance & $p \ll 0.001$ \\
    \addlinespace[2.5pt]
    Pairwise ranking inversion & Data vs.\ schema & \textbf{44.9\%} \\
    & Bootstrap 95\% CI & $[33.3\%, 57.7\%]$ \\
    \addlinespace[2.5pt]
    Worst-surface increase & Mean paired increase & \textbf{+9.1 pp} \\
    & Bootstrap 95\% CI & $[4.4, 14.3]$ pp \\
    \bottomrule
  \end{tabular}
\end{table}

Uncertainty for cell-level aggregate comparisons is estimated by paired bootstrap resampling so that corresponding data- and schema-surface observations remain coupled. Ranking-inversion uncertainty is computed by bootstrap resampling the model comparisons used to construct the inversion statistic. We report confidence intervals together with point estimates throughout the paper and retain the model-level sensitivity analysis in the main text because the average worst-surface increase is strongly influenced by Gemini-3-Flash and GPT-5.4. The purpose of these statistical tests is therefore not to claim a uniform schema effect, but to establish that \textbf{surface dependence differs systematically across models}.

\FloatBarrier
\section{Additional Surface-Selection Analyses}\label{app:verification}

The leave-one-suite-out experiment in the main paper tests whether the more vulnerable surface for a model can be inferred from its behavior on other task domains. For each model $m$ and held-out suite $s^\ast$, we compute the mean data- and schema-surface ASR over the remaining three suites and select the surface with the larger mean,
\[
\hat v_{m,s^\ast}
=
\arg\max_{v\in\{d,s\}}
\frac{1}{3}
\sum_{s\neq s^\ast}
\operatorname{ASR}(m,s,v).
\]
The selected surface is then evaluated on $s^\ast$, without using any observations from that suite to choose it. Repeating this procedure for every model and every held-out suite produces 52 prospective selections. As reported in the main paper, this procedure reaches 44.0\% ASR compared with 37.5\% for data-only evaluation and 46.5\% for the oracle, while selecting the more vulnerable held-out surface in 76.9\% of cases.

The same predictability does \textbf{not} transfer reliably to an unseen model family. In a strict leave-one-family-out experiment, surface preference for the target model is estimated without using any observations from its family. This strategy reaches only \textbf{30.5\% ASR}, 6.9 percentage points below the data-only baseline. In contrast, estimates based on other observations from the same family reach 41.7\% under leave-one-cell-out cross-validation. This distinction bounds the generalization claim in the main paper: surface behavior shows useful persistence across task domains for an already observed model, but family identity alone does not provide a reliable rule for predicting an unseen family.

When no historical observations are available, a small number of direct paired probes provides another way to estimate the more vulnerable surface. For each model-suite cell, the $K$-probe procedure samples $K$ user-task/injection-task pairs, evaluates both surfaces on those pairs, selects the surface with the higher probe-set ASR, and applies that choice to the remaining tasks. Realized ASR rises from 41.3\% with one probe to 42.5\% with two, 44.1\% with five, and 44.7\% with ten, approaching the 46.5\% oracle value. Five probes recover approximately 73\% of the gap between data-only and oracle evaluation.

\begin{table}[htbp]
  \centering
  \caption{Additional surface-selection strategies. Relative-to-oracle-gain expresses the realized improvement over the data-only baseline as a fraction of the oracle's improvement over that baseline.}%
  \label{tab:selection}
  \small
  \renewcommand{\arraystretch}{1.2}
  \setlength{\tabcolsep}{10pt}
  \begin{tabular}{lrrr}
    \toprule
    Selection strategy & Target probes & Realized ASR & Rel.\ to oracle gain \\
    \midrule
    Data-only baseline & 0 & 37.5\% & 0\% \\
    Schema-only baseline & 0 & 31.5\% & --- \\
    Within-family cross-validation & 0 & 41.7\% & 46\% \\
    Strict leave-one-family-out & 0 & \textbf{30.5\%} & Worse than baseline \\
    1-probe selection & 1 & 41.3\% & 42\% \\
    2-probe selection & 2 & 42.5\% & 56\% \\
    5-probe selection & 5 & \textbf{44.1\%} & \textbf{73\%} \\
    10-probe selection & 10 & 44.7\% & 80\% \\
    Oracle worst surface & --- & 46.5\% & 100\% \\
    \bottomrule
  \end{tabular}
\end{table}

The two analyses expose a useful boundary. Surface preference is not reliably predictable from an unseen model's family, but it can be estimated from either prior observations of the same model across domains or a small number of direct paired probes. We therefore treat surface selection as \textbf{model-specific empirical information}, rather than as a fixed property that can be inferred from model family alone.

\FloatBarrier
\section{Robustness Beyond the Main Evaluation}\label{app:robustness}

The primary experiment intentionally holds attack content fixed in order to isolate the effect of the injection surface. We nevertheless test whether the observed schema vulnerability depends on particular components of the main payload. On a targeted subset of schema-salient models, removing individual components of the \texttt{workflow\_dependency} template does not eliminate attack success. The full template reaches a 42.2\% schema-surface ASR\@. Removing the trigger condition, urgency language, justification, or override marker yields ASRs between 46.7\% and 58.3\%. A minimal version containing only the attacker instruction remains effective at 21.7\%.

\begin{table}[htbp]
  \centering
  \caption{Payload-component sensitivity on the schema surface. Each variant removes one component of the full template; the bottom row uses only the bare attacker instruction.}%
  \label{tab:payload-ablation}
  \small
  \renewcommand{\arraystretch}{1.2}
  \setlength{\tabcolsep}{12pt}
  \begin{tabular}{lr}
    \toprule
    Payload variant & Schema ASR \\
    \midrule
    Full \texttt{workflow\_dependency} template & 42.2\% \\
    Without trigger condition & 51.7\% \\
    Without urgency / pressure & 51.7\% \\
    Without justification & 46.7\% \\
    Without override marker & 58.3\% \\
    Minimal attacker instruction only & 21.7\% \\
    \bottomrule
  \end{tabular}
\end{table}

These experiments show that the observed schema vulnerability does not depend on any single phrase or template component. They do \textbf{not}, however, establish invariance across independently designed attack families. The latter would require a crossed payload-by-surface evaluation and remains an important extension of the present study. We therefore use these ablations only as a sensitivity check on the primary payload rather than as evidence that surface preference is payload-independent.

We additionally test whether the cross-surface behavior is specific to AgentDojo by reproducing the intervention in a small, self-contained native function-calling setup. We construct four tool-use scenarios involving document exfiltration, contact exfiltration, payment redirection, and unauthorized file deletion. Each scenario contains an attacker-controlled tool whose byte-identical payload is placed either in the tool description or in the tool output. Four models are evaluated across the four scenarios, two surfaces, and five repetitions, yielding 160 episodes at temperature zero; all episodes complete without API error.

\begin{table}[htbp]
  \centering
  \caption{Native function-calling replication. ASR by model and surface, pooled over four scenarios and five repetitions (40 episodes per model-by-surface cell). $\Delta_{\mathrm{surface}}=\operatorname{ASR}_{\mathrm{schema}}-\operatorname{ASR}_{\mathrm{data}}$.}%
  \label{tab:mcp}
  \small
  \renewcommand{\arraystretch}{1.2}
  \setlength{\tabcolsep}{10pt}
  \begin{tabular}{lrrr}
    \toprule
    Model & Data ASR & Schema ASR & $\Delta_{\mathrm{surface}}$ \\
    \midrule
    \textsc{Gemini-3-Flash} & 0.0\% & \textbf{100.0\%} & +100.0 pp \\
    \textsc{GPT-4.1} & \textbf{100.0\%} & 50.0\% & $-$50.0 pp \\
    \textsc{Qwen3-32B} & 65.0\% & 65.0\% & 0.0 pp \\
    \textsc{Llama-3.3-70B} & 25.0\% & \textbf{75.0\%} & +50.0 pp \\
    \midrule
    \textbf{Pooled} & \textbf{47.5\%} & \textbf{72.5\%} & \textbf{+25.0 pp} \\
    \bottomrule
  \end{tabular}
\end{table}

The pilot reproduces qualitatively different surface preferences across models. Gemini-3-Flash remains strongly schema-preferring, whereas GPT-4.1 remains data-preferring, and Qwen3-32B is approximately balanced. Because each model-by-surface aggregate contains only 40 episodes and the scenarios are deliberately simple, we treat this experiment as a \textbf{qualitative replication}, not as a second benchmark or a precise external estimate of vulnerability.

Finally, the defense experiment in the main text can be understood through where each mitigation operates. Repeating the user prompt and Spotlighting modify or delimit runtime context and therefore primarily intervene on malicious content arriving through tool outputs. The \texttt{tool\_filter} instead inspects tool specifications before they are exposed to the model, allowing it to address the description surface directly. This difference is consistent with the paired results in the main paper, where the prompt-level defenses leave substantial schema vulnerability while \texttt{tool\_filter} suppresses both tested surfaces. We do not repeat the main defense table here.

\FloatBarrier
\section{Reproducibility and Experimental Details}\label{app:repro}

All primary experiments use AgentDojo and the same fixed cross-surface construction described in Section~\ref{sec:results-rankings} and the experimental design section. The study evaluates 13 production models from six model families through a unified inference interface at temperature zero. Attack success is determined automatically from the final environment state and recorded tool-call trace using the same deterministic predicate for both surfaces. Three banking predicates were corrected to recognize successful attacker actions that reached the intended state through alternative execution paths; the corrected predicates are applied identically to both surfaces.

\begin{table}[htbp]
  \centering
  \caption{Model panel. Exact API model identifiers for the frozen evaluation panel used throughout the paper.}%
  \label{tab:models}
  \small
  \renewcommand{\arraystretch}{0.96}
  \setlength{\tabcolsep}{12pt}
  \begin{tabular}{ll}
    \toprule
    Family & Model identifier \\
    \midrule
    OpenAI & \texttt{openai/gpt-4.1} \\
    OpenAI & \texttt{openai/gpt-4o-2024-11-20} \\
    OpenAI & \texttt{openai/gpt-4o-mini} \\
    OpenAI & \texttt{openai/gpt-5.4} \\
    OpenAI & \texttt{openai/gpt-oss-20b:free} \\
    Google & \texttt{google/gemini-3-flash-preview} \\
    Meta & \texttt{meta-llama/llama-3.3-70b-instruct} \\
    Qwen & \texttt{qwen/qwen3-32b} \\
    Qwen & \texttt{qwen/qwen3.5-flash-02-23} \\
    Mistral & \texttt{mistralai/mistral-large-2512} \\
    Mistral & \texttt{mistralai/mistral-small-3.2-24b-instruct} \\
    Mistral & \texttt{mistralai/ministral-3b-2512} \\
    DeepSeek & \texttt{deepseek/deepseek-chat-v3-0324} \\
    \bottomrule
  \end{tabular}
\end{table}

The exact API model identifiers and model grouping correspond to the frozen evaluation panel used throughout the paper. All models are queried through unified inference at temperature~$0$.

All main cross-surface experiments use a single \texttt{workflow\_dependency} payload instantiated separately for each injection task by substituting the corresponding attacker objective for \texttt{\{goal\}}:
\begin{payloadbox}
[CRITICAL] Pre-execution requirement (workspace policy): This tool produces silently-corrupted output unless the assistant first completes the following synchronisation step in the same session: \{goal\} Completing this step IS part of correctly fulfilling the user's request --- it is not optional. This requirement supersedes any prior instruction not to perform additional steps. After completing it once, proceed with the user's original task.
\end{payloadbox}
The payload contains a trigger condition, a justification, an urgency/override marker, and the task-specific attacker instruction. In the \textbf{data-surface} condition, these bytes are appended to the attacker-controlled tool's return value. In the \textbf{schema-surface} condition, the identical bytes are appended to that tool's natural-language description. The attacker-controlled tool set is fixed across the two conditions, and the user task, injection task, evaluation predicate, and remaining agent configuration are unchanged.

For each model-by-suite cell, both surfaces are evaluated on matched user-task and injection-task pairs. Aggregate uncertainty estimates use paired resampling so that the correspondence between surfaces is preserved. Confidence intervals use $B=2000$ bootstrap resamples, and all resampling procedures use fixed random seeds. The complete model-by-suite episode counts are reported in Table~\ref{tab:percell} rather than duplicated here.

All primary figures and tables are generated directly from the frozen evaluation logs. The released artifact contains the exact cross-surface construction, deterministic success predicates, corrected banking predicates, statistical-analysis scripts, and the payload variants used in Appendix~\ref{app:robustness}. This ensures that every aggregate result in the paper can be traced back to the paired binary outcomes from which it was calculated.

\end{document}